\documentclass[journal=jpccck,manuscript=article]{achemso}
\usepackage[version=3]{mhchem} 
\usepackage[T1]{fontenc}       

\usepackage{multirow}  
\usepackage{siunitx} 
\usepackage{xargs} 

\DeclareSIUnit[]\formulaunit{f.u.}

\usepackage{xspace}
\usepackage[colorinlistoftodos,prependcaption,textsize=tiny]{todonotes}
\newcommandx{\change}[2][1=]{\todo[linecolor=blue,backgroundcolor=blue!25,bordercolor=blue,#1]{#2}}

\newcommand{\hf}{HfO$_2$\xspace}
\newcommand{\zr}{ZrO$_2$\xspace}

\newcommand{\sionhf}{$\text{Si}_\text{Hf}$\xspace}
\newcommand{\sioni}{$\text{Si}_\text{I}$\xspace}
\newcommand{\vono}{$\text{V}_\text{O}$\xspace}
\newcommand{\siando}{$\text{Si}_\text{Hf}\text{V}_\text{O}$\xspace}
\newcommand{\donhf}{$\text{D}_\text{Hf}$\xspace}

\DeclareSIUnit[]\formu{f.u.}
\DeclareSIUnit[]\formup{f.u.\%}
\DeclareSIUnit[]\mevformu{meV/f.u.}

\usepackage{nicefrac}
\usepackage{tcolorbox}

\author{Christopher K\"unneth}
\email{kuenneth@hm.edu}

\author{Robin Materlik}%
\author{Max Falkowski}%
\author{Alfred Kersch}%
\email{akersch@hm.edu}
\affiliation{%
	Munich University of Applied Sciences, \\
	Department of Applied Sciences and Mechatronics, \\
	Lothstr. 34, D-80335 Munich
}%

\title[dopedHfO2]{Impact of Four-Valent Doping on the Crystallographic Phase Formation for Ferroelectric \hf from First-Principles: Implications for Ferroelectric Memory and Energy-Related Applications}


\begin{document}

\begin{tocentry}
\begin{figure}[H]
\includegraphics[width=0.35\textwidth]{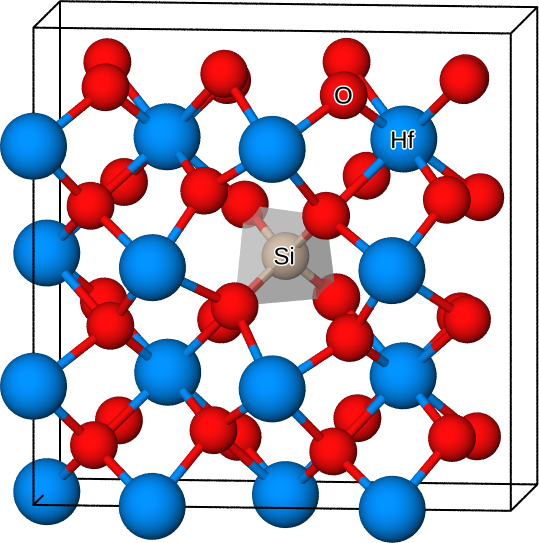}
\caption{}
\end{figure}

\end{tocentry}

\begin{abstract}
The ferroelectric properties of nanoscale silicon doped \hf promise a multitude of applications ranging from ferroelectric memory to energy-related applications. The reason for the unexpected behavior has not been clearly proven and presumably include contributions from size effects and doping effects. Silicon incorporation in \hf is investigated computationally by first-principles using different density functional theory (DFT) methods. Formation energies of interstitial and substitutional silicon in \hf paired with and without an oxygen vacancy prove the substitutional defect as the most likely. Within the investigated concentration window up to 12.5\,formula unit \%, silicon doping alone is not sufficient to stabilize the polar and orthorhombic crystal phase (p-o-phase), which has been identified as the source of the ferroelectricity in \hf. On the other hand, silicon incorporation is one of the strongest promoters of the p-o-phase and the tetragonal phase (t-phase) within the group of investigated dopants, confirming the experimental ferroelectric window. Besides silicon, the favoring effects on the energy of other four-valent dopants, C, Ge, Ti, Sn, Zr and Ce, are examined, revealing Ce as a very promising candidate. The evolution of the volume changes with increasing doping concentration of these four-valent dopants shows an inverse trend for Ce in comparison to silicon. To complement this study, the geometrical incorporation of the dopants in the host \hf lattice was analyzed.
\end{abstract}

\section{Introduction}

In 2011, B\"oscke et al.\cite{Boscke2011} unveiled that silicon doped \hf thin films with a thickness of \SI{10}{nm} exhibit ferroelectricity. Measurements of \SI{2.6}{\formup} silicon doped \hf showed a clear ferroelectric hysteresis. Beginning at \SI{4.3}{\formup} the hysteresis starts to pinch forming a antiferroelectric-like shape. At about \SI{6}{\formup} the ferroelectricity in silicon doped \hf transforms into dielectricity. On the basis of GIXRD measurements, the polar and orthorhombic crystallographic phase Pbc2$_1$ (No. 29, p-o-phase) was proposed as the root of the ferroelectricity\cite{Boscke2011,Sang2015,Muller.2012}. Besides the p-o-phase, other important crystallographic phases could be identified to be present in \hf: (a) the monoclinic P2$_1$/c (No. 14, m-phase), (b) the tetragonal P4$_2$/nmc (No. 137, t-phase), (c) the orthorhombic Pbca (No. 61, o-phase) and (d) the cubic Fm-3m (No. 225, c-phase)\cite{Materlik2015,Sang2015}. Before B\"oscke's finding, the effect of silicon doping on \hf with more than \SI{5}{\formup} was known to stabilize the t-phase and was applied in \SI{50}{nm} HfSiON MIS DRAM trench capacitors\cite{Mueller2005}.

Ferroelectric silicon doped \hf may become of significant technological importance as can be seen in applications like the \SI{28}{nm} FeFET demonstrator\cite{Mikolajick2014}. Nonetheless, its material properties have not been researched satisfactorily. Recently, silicon doped \hf Atomic Layer Deposition (ALD) films for a film thickness of \SI{36}{nm} were explored experimentally in a comprehensive study by Richter et al.\cite{Richter2017} varying the concentration from \SIrange{2.2}{8.3}{\formup}\footnote{In Richter's publication, the ALD cycle ratio is given instead of a dopant concentration because the relative incorporation of Hf and silicon is not known with a high accuracy. Here, we translate the cycle ratio into \si{\formup} assuming equal incorporation rate to be better able to compare computed results with experiments.}. The maximum polarization was found at \SI{4.2}{\formup}. For a higher concentration the hysteresis started to pinch which was interpreted as an increasing t-phase fraction for zero electric field. Contrary, higher electric field can switch the t-phase back to the p-o-phase (field induced ferroelectricity). In addition, a thickness series of \SIrange{5}{60}{nm} with \SI{4.2}{\formup} silicon dopant concentration was prepared showing a maximum polarization around \SI{10}{nm} followed by a decrease and vanishing of remanent polarization at \SI{60}{nm}. In all experiments, the ALD stack was sandwiched between \ce{TiN} electrodes. RevSTM revealed that the crystal phase of the p-o-phase grains at the electrode interface are pinned to the t-phase which implies the existence of a coherent interface. Concluding, the p-o-phase can be stabilized by doping with silicon, but this influence alone is not sufficient. Further mechanism to favour the p-o-phase have been discussed as there are surface and interface energy\cite{Materlik2015,Kunneth2017}, mechanical strain\cite{Materlik2015,ReyesLillo.2014,Batra2017a,Park2014} and electric field\cite{Materlik2015,Batra2017a}.

Besides silicon doping, the p-o-phase in \hf has been stabilized with \ce{Al}, \ce{Sr}, \ce{Y}, \ce{La}, \ce{Gd} and \ce{Zr} but no successful stabilization with the 4-valent dopants \ce{Ti}, \ce{Ge} or \ce{Sn} has been reported. Carbon is contained in ALD films on the level of a few \si{\formup} and its effect as a stabilizer of the t-phase has been emphasized, but the effect on stabilization of the p-o-phase is only indirectly visible\cite{Kim2016}. Other four-valent stabilizers of the p-o-phase have not been reported, although Ge or Ti doped ALD \hf films have been produced finding some t-phase stabilization. From this it appears that silicon is the only four-valent dopant with significant stabilization of the p-o-phase. 

Computationally, the effect of the four-valent dopants C, Si, Ge, Ti, Sn and Zr have been studied by Lee et al.\cite{Lee2008} and Fischer et al.\cite{Fischer2008b}, but only as a stabilizer of the t-phase as the p-o-phase was not known at this time. Lee et al. explained the pronounced tetragonal stabilization with silicon doping with the similarity to the \ce{SiO4} with tetrahedral configuration in quartz which seems energetically favourable. Fischer et al. correlated the energy gain from silicon doping with the ionic radius representing the dopant size in the oxide environment. Furthermore, the absence of the m-phase was explained as an additional size effect \cite{Kunneth2017, Materlik2015}.

The only computational studies to explain the p-o-phase stabilization due to dopants so far are Materlik et. al.\cite{Materlik2017} about the Sr doping investigating a single dopant in detail and an extensive study by Batra et al.\cite{Batra2017b} screening 40 dopants, although omitting small sized dopants like Si, Al and C. Many of those large sized dopants are known from ceramic materials where they are exploited for stabilization of a particular crystallographic phase. Due to the manufacturing process, those ceramic materials are typically larger in grain size than the nanoscaled ALD films and are known to be charge compensated by a accompanying oxygen vacancy. By calculating the formation energy of Sr doped \hf, Materlik et al.\cite{Materlik2017} found that the Sr defect with an associated vacancy does not stabilize the p-o-phase. On the other hand, only Sr doping without vacancy prefers the p-o-phase which is conceivable in a nanoscale Metal-Insulator-Metal (MIM) stack, where the Fermi-level may adjust to reduce the charge occupation and the remaining charge compensation can be provided by interface charges. In accordance with experimental data, Sr was found to stabilize the p-o-phase in a concentration window below \SI{5}{\formup} and the t-phase above that window. However, a destabilization mechanism for the m-phase had to be assumed. Batra et al.\cite{Batra2017b} investigated the stabilization of Ca, Sr, Ba, Y, La and Gd paired with a vacancy in \hf on the crystal phases for 3.125, 6.25 and \SI{12.5}{\formup} doping concentrations concluding that all dopants promote the stabilization of the p-o-phase but doping alone can not stabilize the p-o-phase. 

It is evident that a model for dopant stabilization based solely on monocrystalline properties is incomplete. Care has to be taken to compare computational results with experimental data. As monocrystalline, ferroelectric \hf as such has not been found\footnote{Katayama et al. \cite{Katayama2016} have prepared epitaxial Y doped \hf in the p-o-phase. However, a key to the achievement was the preparation of a ITO bixbyte interlayer.}, the properties of atomic layer deposited (ALD) or chemical solution deposited (CSD) polycrysalline films with grain size in the order of the film thickness are probably closer to a computational investigation than PVD data with much smaller grain size and much higher defect concentration. The comprehensive investigation of PVD prepared doped \hf by Xu et.al. \cite{Xu2017} is therefore out of scope of our investigation.

The purpose of this paper is to close the gap of computational studies on dopants in \hf. First, different DFT methods and its associated energies and volumes for substitutional silicon doping \sionhf are investigated essentially choosing one method. Subsequently, the formation energy and energy differences with respect to the m-phase of a substitutional silicon \sionhf, oxygen vacancy \vono, substitutional silicon paired with a oxygen vacancy \siando and interstitial silicon \sioni are calculated and analyzed. Moreover, we investigate the circumstances to stabilize the p-o-phase of \hf with silicon doping concentration in the known experimental concentration window. Finally, we turn to chemically similar four-valent dopants and perform energetic and structural computations to explore the capability of phase stabilization.

\section{Materials and Methods}
\begin{figure}[tbp]
\includegraphics[width=0.45\textwidth]{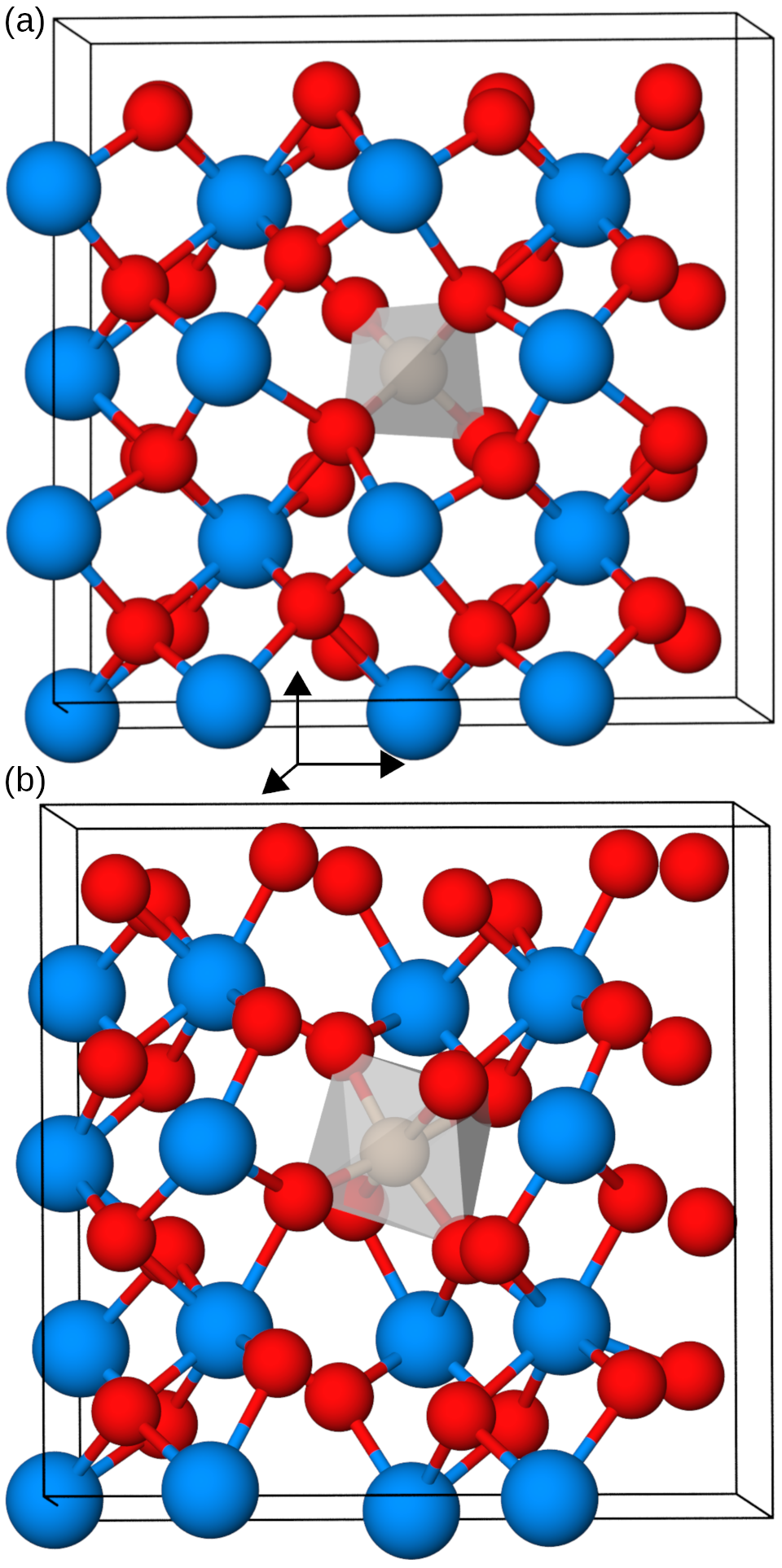}
\caption{\label{fig:bond_orbitals} (a) and (b) exemplifies substitutional defects in \hf (here, \sionhf). Furthermore, (a) illustrates the bonding tetrahedron for the t-phase and (b) the bonding octahedron for the p-o-phase in silicon doped \hf.}
\end{figure}

DFT total energies in this publications were obtained with the (i) all-electron DFT code FHI-Aims \cite{Blum2009,Knuth2015,Marek2014,Auckenthaler2011,Havu2009} which uses numerical atom-centered basis function and (ii) the plane-wave based pseudo-potential code Abinit\cite{Torrent2008,Gonze2016,Gonze2009}. FHI-Aims results were obtained using the Local Density Approximation (AIMS-LDA, PW\cite{Perdew1992} parameterization), Generalized Gradient Approximation (AIMS-PBE, PBE\cite{Perdew1996} approximation) and Heyd-Scuseria-Ernzerhof\cite{Heyd2006,Ren2012a} (AIMS-HSE06) with the mixing parameter $\alpha = 0.25$ and $\omega = 0.11 a_0^{-1}$ for the exchange-correlation (XC) functional. In Abinit only the Local Density Approximation (LDA, PZ\cite{Perdew1981} parameterization) XC functional in combination with projected augmented wave (PAW) pseudo-potentials (PP) from the PP library of Ref. \cite{Garrity2014} (GBRV) were used. The GBRV library contains a Hf PP for the Hf$^{+2}$ and Hf$^{+4}$ ionic configuration referred as GBRV and GBRV*, respectively.

A convergence study reveals that a k-point grid of $6 \times 6 \times 6$ for 12 atoms, $3 \times 6 \times 6$ for 24 atoms, $3 \times 3 \times 6$ for 48 atoms and $3 \times 3 \times 3$ for 96 atoms is sufficient for all FHI-Aims and Abinit calculations with respect to the energies. The electronic (ionic) force was converged until \SI{1e-5}{eV\per\angstrom} (\SI{1e-4}{eV\per\angstrom}) with the tight basis set in the second tier for FHI-Aims and \SI{5e-6}{eV\per\angstrom} (\SI{5e-5}{eV\per\angstrom}) for Abinit. The plane wave and PAW cutoff for the Abinit calculations was \SI{18}{Ha} and \SI{22}{Ha}, respectively. In charged supercells only ions were allowed to move keeping the lattice vectors of the uncharged. Vibrational frequency calculations for the entropy contribution to the free energy were carried out with the utility Phonopy\cite{Togo2015} and Anaddb (included in Abinit) using finite displacements.

Pure \hf calculations for the m-, t- and p-o-phase were carried out in 12 atoms and for the o-phase in 24 atoms sized unit cells. \SI{6.25}{\formup} ($\text{f.u} = \nicefrac{n}{3}$ with $n$ the number of atoms) doping was achieved by substituting one Hf with a dopant \donhf in a 48 atoms sized unit cell which is exemplified for silicon in FIG. \ref{fig:bond_orbitals}. Since the 48 atoms sized unit cell can be created expanding the 12 atoms sized unit cell in the three distinct directions for the m-, t- and f-phase, all three choices were calculated and the lowest energy was chosen. Consequently, in the case of the o-phase, the 24 atoms sized unit cell was expanded in two directions and, again, the lowest energy was chosen. In contrast, for \SI{3.125}{\formup} doping the supercell was uniquely built with the multiplication of $2 \times 2 \times 2$ of the 12 atoms sized cells and $2 \times 2 \times 1$ of the 24 atoms unit cells. Doping concentrations in this publication are specified in \si{\formup} which is in the case of metal substitution the same as \si{cat.\%} but differs from \si{ani.\%}. As anion and cation doping is used simultaneously in graphs, \si{\formup} is used instead throughout the paper. Since FHI-Aims does not include symmetry considerations, all convergences were archived without symmetry constraints. To find the preferred oxygen vacancy positions in silicon doped and pure \hf, the energy of all symmetry inequivalent positions was calculated. Finally, the vacancy position of the lowest energy was chosen. The final lattice constants and band gap are tabulated in the Supporting Information.

The formation energy $E^\zeta_\text{f}$ for a phase $\zeta \in \left\{\text{m, o, p-o, t}\right\}$ are calculated according to \cite{Freysoldt2014}
\begin{eqnarray}
E_\text{f}^\zeta\left[X^q\right] &&= E_\text{tot}^\zeta\left[X^q\right] - E_\text{tot}^\zeta\left[\text{pure}^0\right] -\sum_i n_i \mu_i \nonumber\\ 
&& + q\left( E_\text{F} + E_\text{VBM}^\zeta\left[\text{pure}^0\right] + \Delta V^\zeta\left[X^0\right] \right) \nonumber\\
&&+ E_\text{corr}^\zeta[X^q]
\label{eq:formation_energy}
\end{eqnarray}

with $E^\zeta_\text{tot}$ the total energy of phase $\zeta$, $n_i$ the numbers of impurities, $\mu_i$ the chemical potential of the impurity $i$, $E_\text{F}$ the Fermi level referenced to the energy of the valence band maximum $E^\zeta_\text{VBM}$, $\Delta V^\zeta$ the potential alignment, $E^\zeta_\text{corr}$ the charge correction due to finite size of the unit cell and $ X \in \left \{\text{V}_\text{O}, \text{Si}_\text{Hf}, \text{Si}_\text{Hf}\text{V}_\text{O}\right\}$ the defect. Calculations for charged structures were carried out for the charges $q=-3,\ldots,+3$ for all three defects with the lattice fixed to the uncharged structure. For $ X \in \left \{ \text{Si}_\text{i}, \text{C}_\text{Hf}, \text{Ge}_\text{Hf},\text{Ti}_\text{Hf},\text{Sn}_\text{Hf},\text{Zr}_\text{Hf},\text{Ce}_\text{Hf} \right\}$ only calculation for charge $q=0$ were carried out.

The chemical potentials of \vono, \sionhf and \siando were $\sum_i n_i \mu_i = -\mu_\text{O}$, $\sum_i n_i \mu_i = -\mu_\text{Si} + \mu_\text{Hf}$ and $\sum_i n_i \mu_i = -\mu_\text{Si} + \mu_\text{Hf} -\mu_\text{O}$, respectively. Under oxygen deficient conditions $\mu_\text{O}$ is calculated from \ce{TiO2} ($\mu_\text{O}^{\text{TiO}_2}$) in anatase structure and under oxygen rich from \ce{O2} ($\mu_\text{O}^{\text{O}_2}$). $\mu_\text{Hf}$ was calculated using $\alpha$-Hf. The chemical potentials $\mu_\text{C}$, $\mu_\text{Ge}$, $\mu_\text{Ti}$, $\mu_\text{Sn}$, $\mu_\text{Zr}$ and $\mu_\text{Ce}$ were calculated from diamond C, diamond Ge, hexagonal (P6$_3$/mmc, No. 194) Ti, beta Sn, hexagonal (P6$_3$/mmc, No. 194) Zr and cubic (Fm$\bar{3}$m, No. 225), respectively.  Figures of the atomic structures in this publication are produced with Ovito\cite{Stukowski2010}. If $q$ is omitted in the notation, the charge is set $0$.

\section{Results}
\subsection{Si doping with different DFT methods}
\begin{figure}[H]
\includegraphics[width=0.5\textwidth]{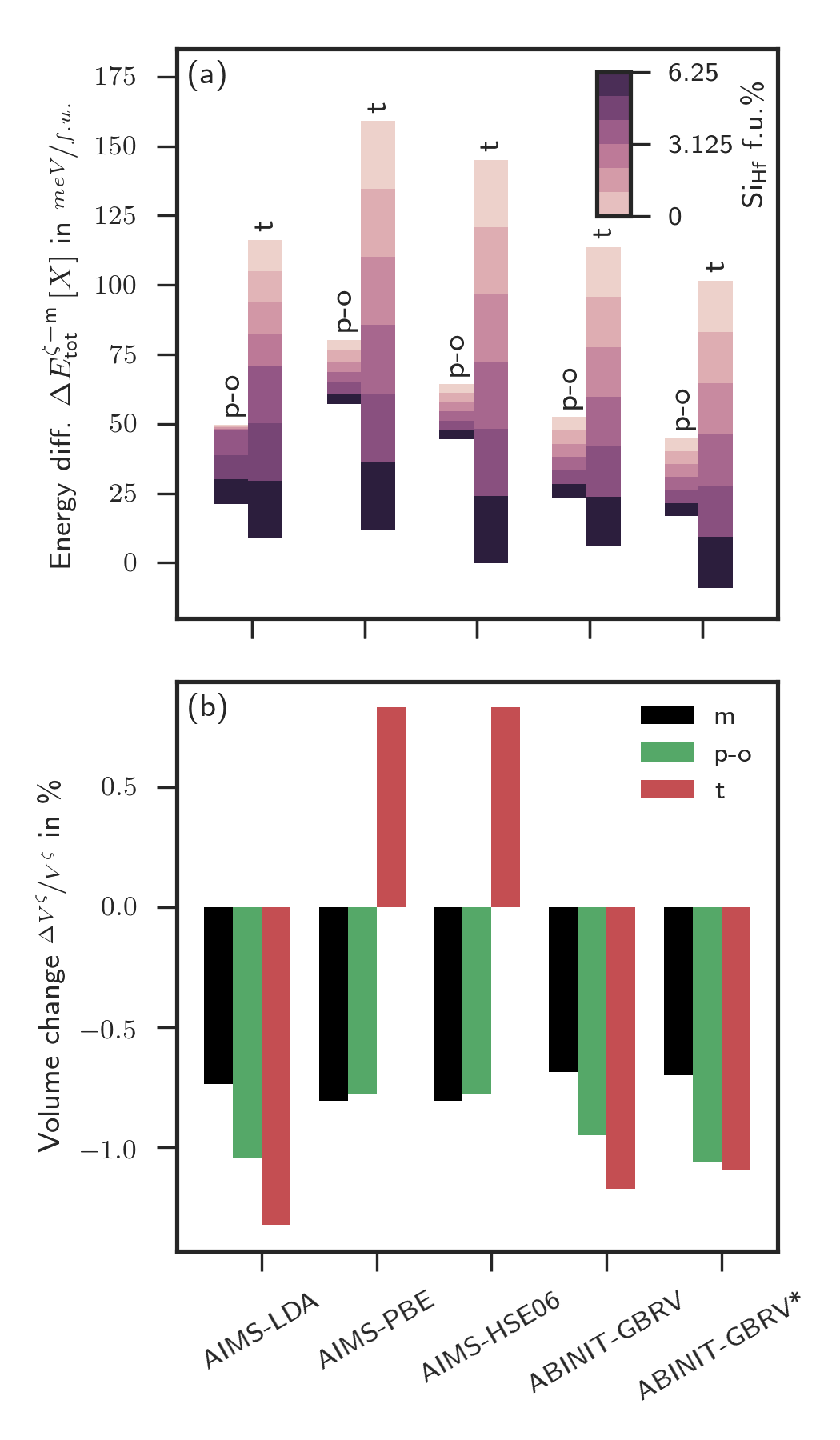}
\caption{\label{fig:energy_different_methods} The energy differences with respect to the m-phase for five different DFT methods up to a doping concentration of \SI{6.25}{\formup} for silicon doped \hf are illustrated in (a). The associated volume change with respect to undoped \hf is shown in (b). Except for AIMS-LDA, the values between pure and \SI{6.25}{\formup} doping concentration were linear interpolated. In the case of AIMS-LDA, between 0 and \SI{3.125}{\formup}, and 3.125 and \SI{6.25}{\formup} was interpolated. The coordinates and lattice parameters for HSE06 calculations was fixed to PBE.}
\end{figure}

To choose a consistent basis for further calculations, different DFT methods with different XC functionals were first evaluated. FIG. \ref{fig:energy_different_methods} compiles the energy differences (a) with respect to the m-phase and the associated volume changes (b) for the five chosen DFT methods. All methods shows consistently the favouring of the p-o- and t-phase with increasing silicon concentration. Between 4 and \SI{6}{\formup}, the t-phase becomes the most stable for all methods. Moreover, no method shows the p-o-phase to be the lowest in energy for any concentration. Therefore, silicon doping is not a mechanism to exclude neither the m- nor o-phase as the thermodynamic most and second most favourable crystal structures of the monocrystalline material respectively. Besides doping, a high negative entropic influence on the energy differences of the t-phase and a smaller on the p-o-phase from temperature is expected. Entropy contributions from phonon modes are partly listed in TAB. \ref{tab:stabilization} for $T=\SI{300}{K}$. However, the additional energy contributions from entropy do not alter the energy picture in general leaving the trends unaffected. All further calculations in this publication are carried out using AIMS-LDA.

In addition to the energy effect, silicon incorporation causes a change of the volume. The volumes for all DFT methods in FIG. \ref{fig:energy_different_methods} (b) are decreasing after silicon installation. Only the trend of the t-phase with AIMS-PBE is to the opposite direction. Since silicon is smaller than Hf, a decreasing volume is believed to be the more reasonable trend. Although the t-phase data point was carefully checked, no error in the calculation and the analysis could be found. Experimentally, Zhao et al.\cite{Zhao2014} precisely measured the volume change by silicon doping in \hf ceramics. Interestingly, in this study, only the m-phase was found up to a doping limit of \SI{9}{\formup} silicon accompanied with no significant change in the unit cell volume. 

FIG. \ref{fig:energy_different_methods} discusses the question, how capable currently used XC functionals are in reflecting the crystallographic phase stability which requires relative total energy values on the level of a few \si{meV/\formu}. For undoped \hf and \zr several comparisons between LDA and PBE XC functionals\cite{Lowther1999,Fadda2010} (local) and also more recently hybrid XC functionals\cite{Barabash2017} (non-local) were carried out. The results show generally larger energy differences between the phases for PBE than LDA but maintaining the energetic order and yield no contradiction with structural data. HSE06 hybrid functional calculations from Barabash et al.\cite{Barabash2017} give values energetically between PBE and LDA but closer to PBE which is similar to our calculations. Total energy differences have been studied with HSE06 in \ce{TiO2}\cite{Arroyo-DeDompablo2011}. Although, the structural results were superior with the local functionals, the anatase phase turned out to be lower than rutile which contradicts the experiment. Either the ground state is obscured from further effects, similar to polycristalline \hf, or better total energy results are not guaranteed with the HSE06 functional as the fraction of the exact exchange in the method is optimized to match the band gap. AIMS-HSE06 in FIG. \ref{fig:energy_different_methods} are single point calculations using the coordinates and lattice constants of AIMS-PBE. 

\begin{table}[tb]
\caption{\label{tab:stabilization} Energy differences relative to m-phase and volume changes relative to the undoped structure for different used DFT methods are presented. Silicon and vacancy doping are both for \SI{6.25}{\formup}. Values in parentheses are the energies including the vibrational entropy contribution from phonon modes for $T = \SI{300}{K}$. $\Delta E_\text{tot}^{\zeta-\text{m}} = E_\text{tot}^{\zeta}\left[X\right] - E_\text{tot}^{\text{m}}\left[X\right] $ and $\nicefrac{\Delta V^\zeta}{V^\zeta} = \nicefrac{\left( V^{\zeta}[X] - V^\zeta[\text{pure}] \right)}{V^{\zeta}[\text{pure}] }$.}
\begin{tabular}{ll|ccc|cccc}
\hline
 & & \multicolumn{3}{c|}{$\Delta E_\text{tot}^{\zeta-\text{m}}$}  & \multicolumn{4}{c}{$\nicefrac{\Delta V^\zeta}{V^\zeta}$} \\
	$X$     &                        &       o        &      p-o       &       t        &   m   &   o   &  p-o  &   t     \\
	        &                        & \si{\mevformu} & \si{\mevformu} & \si{\mevformu} &  \%   &  \%   &  \%   &  \%     \\ \hline
	pure    &  \multirow{2}{*}{AIMS-}  &      28.1 (27.7)      &      49.5 (48.8)  &  115.8 (99.4) &   0   &   0   &   0   &   0     \\
	\sionhf &                        &      18.5 (16.4)    &      21.1 (19.7) &   8.7 (-0.8)  & -0.74 & -1.15 & -1.04 & -1.33   \\
	\vono   &  \multirow{3}{*}{LDA}  &      30.0      &      40.1      &     104.0      & 1.74  & 1.87  & 1.71  & 1.71    \\
	\siando &                        &      13.5      &      25.5      &      9.1       & 2.80  & 1.61  & -0.22 & 0.85    \\ 
	\sioni  &                        &      12.4      &      109.8     &      93.8      & 2.15  & -0.61 & 2.43  & 4.32    \\ \hline
	pure    &  AIMS-                 &      28.4      & 80.0 (78.9)    &     158.9 (135.5)      &   0   &   0   &   0   &   0     \\
	\sionhf &  PBE                   &      22.3      &      57.0      &      0.0       & -0.81 &   -0.96   & -0.78 & 0.83     \\ \hline
	pure    & AIMS-                  &       -        &      64.0      &     145.0      &   0   &   0   &   0   &   0     \\
	\sionhf & HSE06                  &       -        &      44.4      &      0.0       & -0.81 &   -   & -0.78 & 0.83    \\ \hline
	pure    & \multirow{2}{*}{GBRV}  &      27.0      &      52.2 (52.2)      &     113.6 (93.0)     &   0   &   0   &   0   &   0     \\
	\sionhf &                        &       -        &      23.4      &      6.0       & -0.69 &   -   & -0.95 & -1.18   \\ \hline
	pure    & \multirow{2}{*}{GBRV*} &       -        &      44.5      &     101.4      &   0   &   0   &   0   &   0     \\
	\sionhf &                        &       -        &      16.8      &      -9.0      & -0.70 &   -   & -1.06 & -1.10   \\ \hline
\end{tabular}
\end{table}

\subsection{Formation energy of Si related defects}

\begin{figure}[tb]
\includegraphics{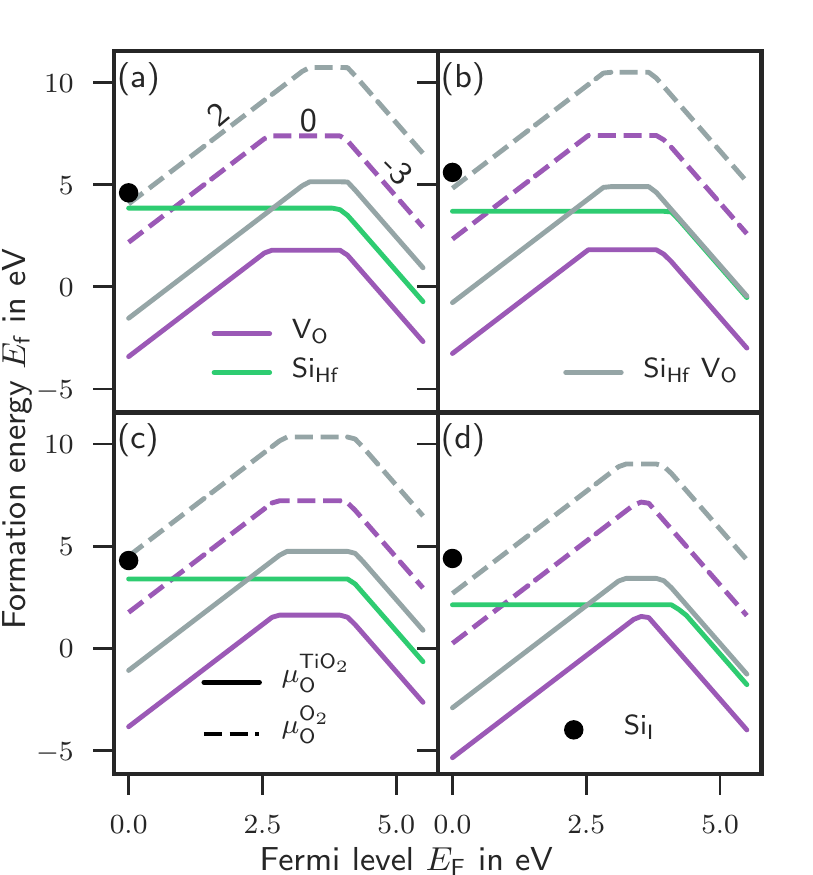}
\caption{\label{fig:formation_energies} The subplots (a), (b), (c) and (d) show the formation energy for \vono, \sionhf and \siando over the Fermi level for the m-, o-, p-o- and t-phase, respectively. The used energies were taken from \SI{6.25}{\formup} doping. The small numbers indicate the charge state $q$ of the defect and the black dots the formation energy of interstitial silicon \sioni.}
\end{figure}

The substitutional defect structure \sionhf is created most likely when silicon is incorporated in the \hf lattice. To substantiate this statement we have calculated the formation energy for the \sionhf, \siando, \vono and \sioni defect in all the crystal phases for two oxygen partial pressures, shown in FIG. \ref{fig:formation_energies}. For oxygen deficient (poor) conditions, the chemical potential with \ce{TiO2} (solid lines)  was used and for oxygen rich conditions with \ce{O2} (dashed lines). \ce{TiO2} was chosen since it corresponds to a typical, oxidized electrode material in \hf thin films and \ce{O2} is a typical precursor in the ALD process for such films. In FIG. \ref{fig:formation_energies}, only the charge state $q$ with the lowest formation energy is depicted. Therefore, the kinks indicate the thermodynamic charge transitions levels. 

A comparison of the oxygen deficient case with the oxygen rich for all subplots in FIG. \ref{fig:formation_energies} shows that the formation energies of \vono and \siando are shifted by a constant value of $\mu_\text{O}^{\text{TiO}_2} - \mu_\text{O}^{\text{O}_2} = \SI{5.6}{eV}$ leaving the \sionhf unaffected. In the oxygen rich case ($\mu_\text{O}^{\text{O}_2}$), the formation of the \vono and \siando are both unfavorable for the Fermi level higher than \SI{\approx 0.7}{eV} in comparison with \sionhf for all phases. Assuming that the $\mu_\text{O}^{\text{O}_2}$ chemical potential is close to the production conditions of the thin films, silicon doping preferentially creates \sionhf. 

After the production process, the oxygen partial pressure is determined by $\mu_\text{O}^{\text{TiO}_2}$ favouring the creation of \vono and \siando defects with the necessary formation energy dependent on the Fermi level. The newly created and mobile \vono defects can recombine with the already present immobile \sionhf defects to \siando releasing an energy 0.5, 0.6, 0.28 and \SI{0.27}{eV} using solely formation energies of $q=0$ for the m-, o-, p-o- and t-phase, respectively, for the reaction \sionhf + \vono $\rightarrow$ \siando. In comparison, the energy release for the analogous reaction in Sr doped \hf is \SI{2.4}{eV} for the p-o-phase which is approximately an order of magnitude higher than for silicon doped \hf. 

Besides for the \sionhf, \vono and \siando defects, formation energies for silicon interstitials \sioni were carried out only for the charge $q=0$. Placing the \sioni in all symmetry inequivalent polyhedrons spanned by adjacent atoms for each of crystal phase the lowest \sioni formation energies were found to be 4.6, 5.6, 4.3 and \SI{4.4}{eV} for the m-, o-, p-o- and t-phase, respectively, indicated by the black dots in FIG. \ref{fig:formation_energies}. Consistently, the formation energies of \sioni are higher than \sionhf making them more unlikely.

In addition, the subplots of FIG. \ref{fig:formation_energies} evince charge transition levels at approximately the same Fermi levels for all phases. Except for \sionhf all lines have two transition levels indicating that \sionhf introduces a transition level in a distance of approximately \SI{4}{eV} from the valence band edge. It should be noted, that those levels are close to the conduction band edge predicted by LDA and the remaining difference can be due to uncertainties of the chosen DFT XC functional. The same arguments hold for the \vono defect, which also introduces a level very close to the level of \sionhf at about \SI{4}{eV} with respect to the valence band edge. The band gaps of the calculations can be found in the Supporting Information. Since the deep charge transition level from +2 to 0 at about \SI{2.5}{eV} is only present for \siando and \vono, the level must be introduced by the vacancy. 

We conclude that the substitutional incorporation of silicon \sionhf is favoured for all phases. Those defects are uncharged and do not introduce defect levels in the band gap. Next likely is the creation of oxygen vacancies \vono under operating condition. This defect may combine with \sionhf to create \siando.

\begin{figure}[tb]
\includegraphics{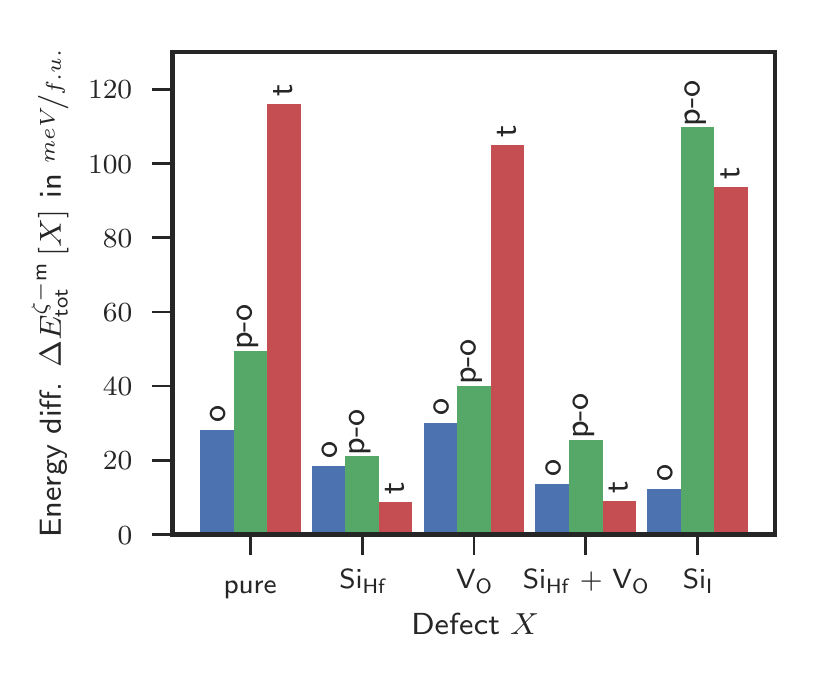}
\caption{\label{fig:energy_furtherdefects} Shows the total energy differences to the m-phase for no defect (pure), \sionhf, \siando, \vono and \sioni for a silicon or vacancy concentration of \SI{6.25}{\formup}.}
\end{figure}

The impact of the discussed defects on the phase stability for \SI{6.25}{\formup} is shown in Figure \ref{fig:energy_furtherdefects}. The stabilization of the t- and p-o-phase with \sionhf is identical with the values shown in FIG. \ref{fig:energy_different_methods} (a) for AIMS-LDA. The vacancy \vono introduces a small stabilization effect which can be neglected in comparison to the \sionhf defect. The energy change of the phases due to incorporation of \siando almost matches the magnitude of \sionhf. On the other hand, silicon interstitials \sioni promote the destabilization of the p-o-phase and a slightly stabilization of the t-phase. Altogether, the phase stabilization is affected by silicon related defects in \hf but up to \SI{6.25}{\formup} the p-o-phase is not shifted to the ground state.

\subsection{Si doping concentration}

On the basis of the formation energies it was concluded that the \sionhf defect is the most likely. We now focus on the impact of \sionhf on the phase stability depending on its concentration. Different doping concentrations were modeled by substituting one metal with one silicon for different sized supercells. All the metal positions in our crystallographic phases are symmetry equivalent. Substituting of one atom out of 96 atoms gives \SI{3.125}{\formup}, one out of 48 gives \SI{6.25}{\formup} and one out of 12 gives \SI{12.5}{\formup}. 

The supercell of 48 atoms can be created by duplicating the 12 atomic unit cell by $2 \times 2 \times 1$, $2 \times 1 \times 2$ and $1 \times 2 \times 2$ except for the o-phase (smallest unit cell has 24 atoms) at which only two meaningful directions are available. Since the energies of these structures showed a significant difference, the structures with the lowest energy for all phases were selected. The c-phase proved to be unstable in all doping concentrations and supercells and, therefore, is excluded in the discussion. 

\begin{figure}[tb]
\includegraphics{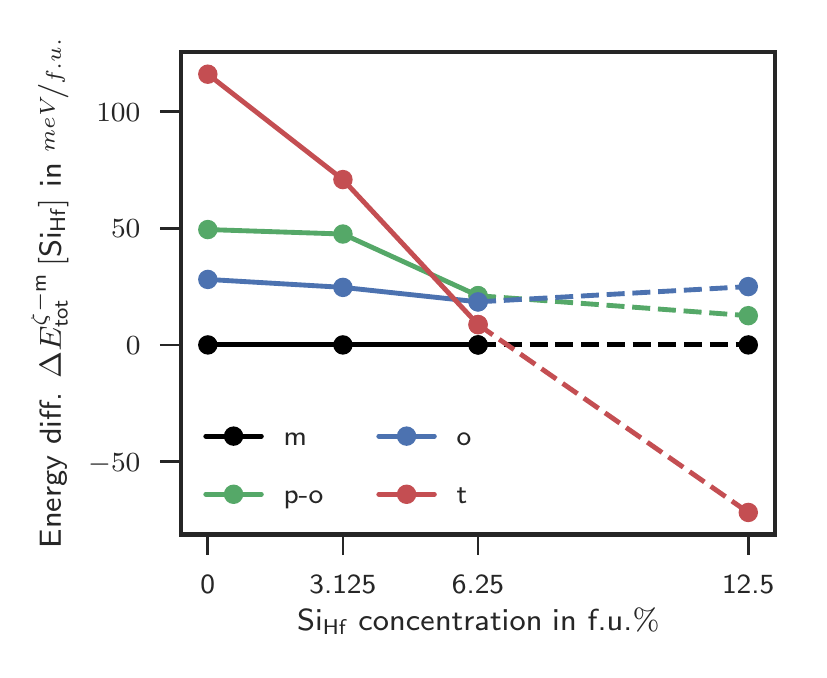}
\caption{\label{fig:energy_concentration} Shows the energy difference with respect to the m-phase for different \sionhf concentrations. The dashed lines indicate that the atom positions and the lattice parameters of these calculations show a significant difference to the actual phase.}
\end{figure}

The energies for all phases for three distinct silicon doping concentrations displayed in FIG. \ref{fig:energy_concentration}, clearly show the t-phase as the ground state for a doping concentration larger than \SI{7}{\formup}. Assuming the m-phase is eliminated by the size effect as discussed previously, the transition to the t-phase is determined from the intersection with the p-o-phase at around \SI{5.7}{\formup}. For lower concentration, the phase with the lowest energy is the high pressure o-phase. In general, to achieve a stabilization in a particular concentration window of the p-o-phase, we either have to assume a destabilization mechanism for the o-phase similar to the m-phase or the phase transformations must be prevented due to a high barrier. Furthermore, the o-phase energy difference seems unaffected by silicon doping. 

A further result concerns the linearity of the energy with the silicon concentration which is obviously not fully realized, especially for the p-o-phase. Due to periodic boundary conditions, the 48 atomic supercells require one crystallographic axis where the silicon atoms are closer to each other than in the other directions. Another supercell to model \SI{6.25}{\formup} doping would be substituting two atoms out of 96 atoms. Such supercell would enable the modelling of the silicon to silicon attraction and repulsion, and their influence on the total energy. However, the systematic investigation of the silicon-silicon (or more general dopant-dopant) interaction is computationally very time consuming. We will report about this effect in a further publication. Regarding these nonlinear effects, the results in FIG. \ref{fig:energy_concentration} for doping concentration larger than \SI{6.25}{\formup} should be interpreted carefully.

\subsection{\label{subsec:four-valent}Other four-valent dopants}
\begin{figure}[tb]
\includegraphics{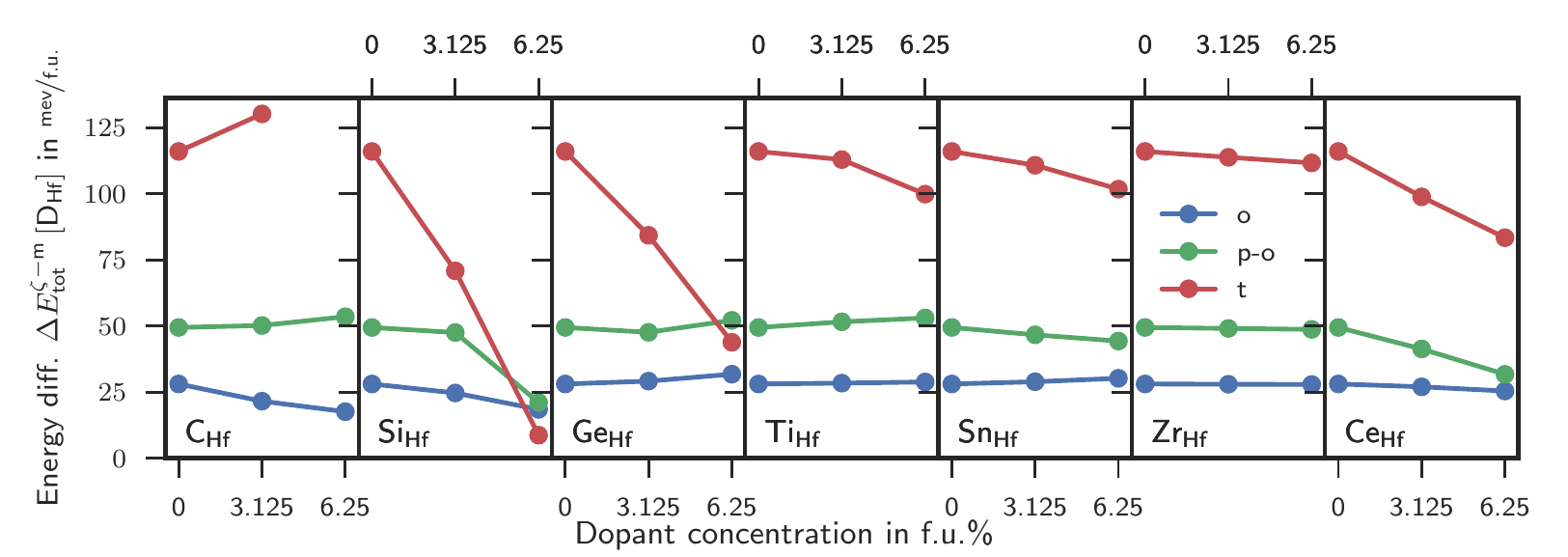}
\caption{\label{fig:energies_4valent} Displays the evolution of the energy differences to the m-phase with increasing doping concentration.}
\end{figure}

After studying the effect of the silicon related defects and doping concentration on the crystallographic phase formation as a prototype system, the effects of other four-valent dopants D are elaborated. C, Ge, and Sn from the carbon group and Ti, and Zr from the titanium group are selected. Furthermore Ce from the Lanthanides because it has a stable $+4$ oxidation state. Motivated from the analysis of the silicon defect, we limited our investigation to substitutional defects on the hafnium site \donhf.

FIG. \ref{fig:energies_4valent} shows the evolution of the energy differences with respect to the m-phase from pure \hf to \SI{6.25}{\formup} doping concentration. Although a favouring effect for some dopants on the p-o-phase is evident, none of the investigated dopants alone shifts the p-o-phase to the lowest energy. For a \hf thin film exhibiting ferroelectricity, a destabilization mechanism for the m- and o-phase has to be assumed promoting the p-o-phase to the lowest in energy. Possible destabilization mechanism have been discussed in literature in Refs. \cite{Materlik2015,Batra2015,Kunneth2017}. Comparing the evolution of the o-phase in all seven subplots in FIG. \ref{fig:energies_4valent}, it can be concluded to be insensitive to doping. Still little sensitive to doping but responding in the case of Si and Ce is the p-o-phase. By contrast, the t-phase sensitivity is high especially in the case of C, Si, Ge and Ce. 

Several attempts were made to find some general geometrical argument for the sensitivity of the crystal phase on the dopant properties. However, none of the investigated correlations between the energy difference, volume, Shannon radius or coordination number of the polyhedron of the dopants rise to the level of causation in the view of the authors. Nevertheless, these correlations can be found in the Supporting Information. Since a causation of the energy and a geometrical property of the dopant was not found it can be concluded that the major effect on the energy differences is of chemical nature. 

\begin{table}[tb]
\caption{\label{tab:volume_abs_diff_energy_dopants} Shows the volume change $\nicefrac{\Delta V^\zeta}{V^\zeta}$ with respect to the host crystal  and the energy difference $\Delta E_\text{tot}^{\zeta-\text{m}}[\text{D}_\text{Hf}-\text{pure}]$ with respect to the undoped m-phase of the defects \donhf for \SI{3.125}{\formup} (\SI{6.25}{\formup}) doping. The crystal radius r$_\text{c}$ is for the coordination number of the t-phase according to\cite{Shannon1976}.}
\begin{tabular}{cc|cccc|ccc}

D & r$_\text{c}$& \multicolumn{4}{c|}{$\nicefrac{\Delta V^\zeta}{V^\zeta}$} & \multicolumn{3}{c}{$\Delta E_\text{tot}^{\zeta-\text{m}}[\text{D}_\text{Hf}-\text{pure}]$}\\
 &  &  m & o & p-o & t & o & p-o & t \\
& pm  & \% & \% & \% & \% & \si{\mevformu} & \si{\mevformu} & \si{\mevformu}  \\

\hline
C  &    29 &    0.3 (0.2) &    0.2 (0.1) &    0.2 (0.1) &  -0.8 (-) &  -6.5 (-10.5) &     0.7 (4.1) &    14.2 (-) \\
Si &    40 &  -0.8 (-1.0) &  -0.9 (-1.5) &  -0.5 (-1.3) &  -0.6 (-1.7) &   -3.4 (-9.6) &  -1.9 (-28.3) &  -45.2 (-107.3) \\
Ge &    53 &  -0.4 (-0.8) &  -0.7 (-1.3) &  -0.2 (-0.3) &  -0.4 (-1.0) &     1.1 (3.7) &    -1.8 (2.7) &   -31.7 (-72.2) \\
Ti &    56 &  -0.6 (-1.3) &  -0.7 (-1.4) &  -0.6 (-1.2) &  -0.3 (-0.8) &     0.3 (0.8) &     2.1 (3.6) &    -3.1 (-16.2) \\
Sn &    69 &    0.1 (0.2) &    0.1 (0.1) &    0.1 (0.2) &    0.2 (0.3) &     0.8 (2.1) &   -2.8 (-5.1) &    -5.2 (-14.2) \\
Zr &    98 &    0.1 (0.2) &    0.1 (0.2) &    0.1 (0.2) &    0.0 (0.1) &   -0.1 (-0.2) &   -0.4 (-0.7) &     -2.2 (-4.3) \\
Ce &    111 &    1.0 (1.9) &    1.0 (2.0) &    1.0 (1.9) &    0.8 (1.6) &   -1.1 (-2.7) &  -8.1 (-17.7) &   -17.2 (-32.6) \\
\hline
\end{tabular}
\end{table}

TAB. \ref{tab:volume_abs_diff_energy_dopants} collects the results of the energy difference between $\zeta$-phase and m-phase with respect to the undoped energy difference as 
\begin{eqnarray}
\Delta E_\text{tot}^{\zeta-\text{m}}[\text{D}_\text{Hf}-\text{pure}] && = \Delta E_\text{tot}^{\zeta-\text{m}}[\text{D}_\text{Hf}] - \Delta E_\text{tot}^{\zeta-\text{m}}[\text{pure}] \nonumber\\
&&= \left( E_\text{tot}^{\zeta}[\text{D}_\text{Hf}] - E_\text{tot}^{\text{m}}[\text{D}_\text{Hf}] \right) - \left( E_\text{tot}^{\zeta}[\text{pure}] - E_\text{tot}^{\text{m}}[\text{pure}] \right) \; \text{.}
\end{eqnarray}
Negative values of $\Delta E_\text{tot}^{\zeta-\text{m}}[\text{D}_\text{Hf}-\text{pure}]$ stabilize and positive destabilize the corresponding crystal phase due to doping. Silicon with a value of \SI{-28.3}{\mevformu} at \SI{6.25}{\formup} is by far the best facilitator of the p-o-phase in FIG. \ref{fig:energies_4valent}, but simultaneously the t-phase is preferred by \SI{-107.3}{\mevformu} causing the narrow ferroelectric concentration window observed in experiments. Besides silicon, Ce on the second rank favours the p-o-phase by \SI{-17.7}{\mevformu} and the t-phase by \SI{-32.6}{\mevformu} with a much better p-o- to t-phase ratio of 0.54 in comparison to silicon with 0.26, opening possibly a wide concentration window for the p-o-phase. Sn has a similar, but much smaller capability to favour the p-o-phase and t-phase. The marginal support of Zr for the p-o-phase is amplified by the excellent solubility in Hf up to pure ZrO$_2$. C, Ge and Ti do not support the p-o-phase, but only the t-phase.

\begin{figure}[tb]
\includegraphics{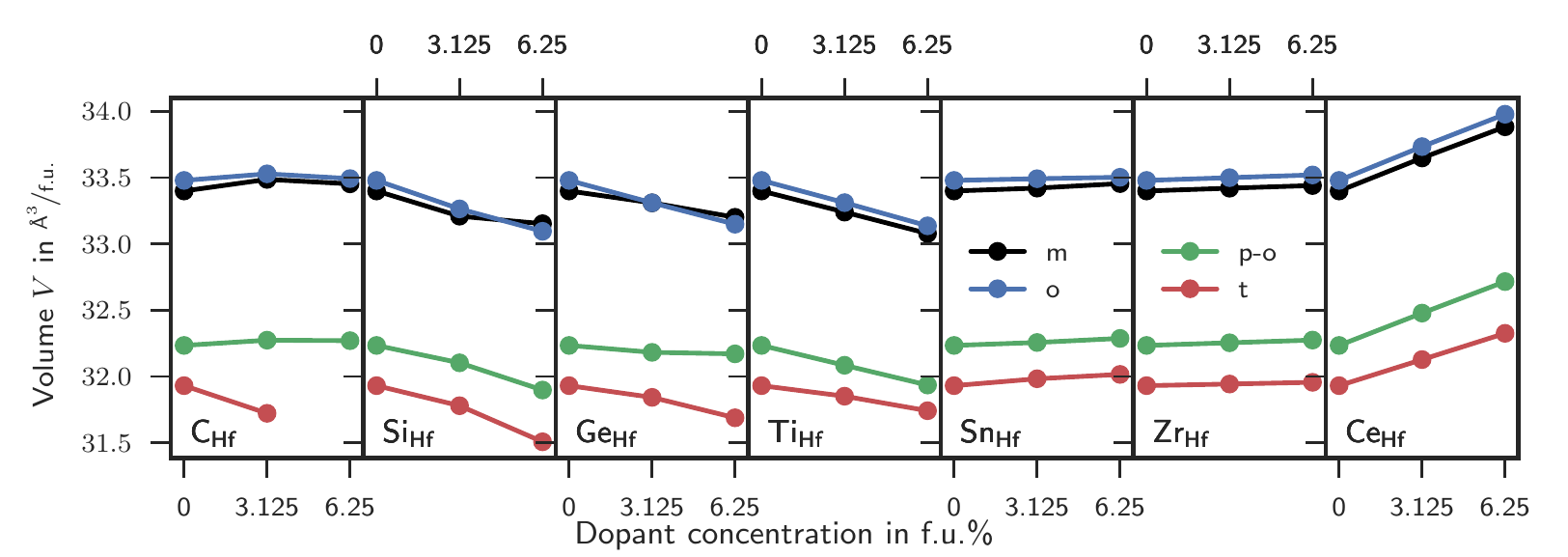}
\caption{\label{fig:volumes_4valent} Shows the evolution of the volumes with increasing doping concentration.}
\end{figure}

The track of the volume change of the four-valent dopants with increasing concentration is illustrated in FIG. \ref{fig:volumes_4valent}. First of all, the trend of all crystal phases for each dopant expose the same volume evolution with increasing dopant concentration. This unit cell volume evolution was correlated with the Shannon radii of Ref. \cite{Shannon1976}, but no simple relation could be found for both small and large ions. For only large ions, the volume increases with ion radius (see Supporting Information). The m- and o-phase have almost the same absolute volume which is about \SI{5}{\%} bigger than the volume of the p-o- and t-phase. Furthermore, the volumes are systematically smaller than experimentally measured volumes confirming the LDA paradigm to always predict smaller volumes. Following the argumentation of Clima et al. \cite{Clima2014} that the volume of the dopant is inversely proportional to the coercive field necessary for ferroelectric switching, Ce doping has the lowest and silicon the highest coercive field of this set of dopants. Volume changes with respect to the undoped phase for the different dopants are provided in TAB. \ref{tab:volume_abs_diff_energy_dopants}

\begin{figure}[tb]
\includegraphics{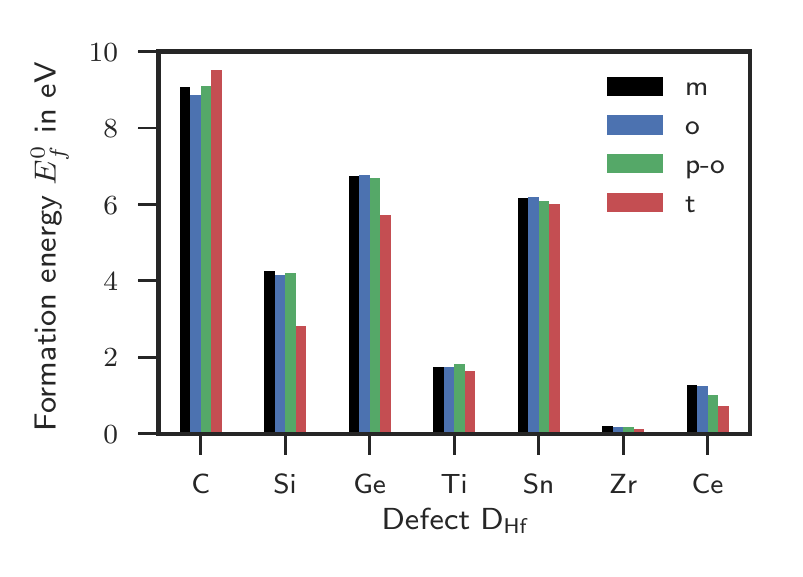}
\caption{\label{fig:formation_energy_4-valent}  The formation energy for charge $q=0$ for \SI{3.125}{\formup} doping.}
\end{figure}

The volume change provides an estimate of the dopant stress exerted to the host crystal. The large volume change of silicon indicates a large force from silicon on the host lattice. For the smaller carbon, the binding in the host crystal is incomplete with a smaller forces and less volume change. The arrangement is chosen from left to right in increasing crystal radii according to Ref. \cite{Shannon1976}. Apparently, a general trend in FIG. \ref{fig:volumes_4valent} is that with increasing radii the volume switch from decreasing to increasing crossing zero between Ti and Sn which is close the radius of Hf with \SI{83}{pm} according to Ref. \cite{Shannon1976}. Although an energy difference to volume correlation is suggested by FIG. \ref{fig:energies_4valent} and \ref{fig:volumes_4valent}, no generally valid relation could be found. However, we included the correlation in the Supporting Information. It has to be taken in mind that the deformation energy of the host crystal, calculated from the volume change and modulus of compressibility, is in the order of \SI{1}{\mevformu} and therefore only a fraction of the energy introduced into the system with the energy of formation.

The formation energies for $q=0$ charged, \SI{3.125}{\formup} doped unit cells are compiled in a bar plot in FIG. \ref{fig:formation_energy_4-valent}. The energies were calculated using EQ. \ref{eq:formation_energy} using the chemical potentials from the metals for $q=0$ and are tabulated in TAB. \ref{tab:formation_dopants}. The dopants are arranged in the same increasing order of crystal radii as in the previous figures and tables which immediately demonstrate that the relation between the dopant size and defect formation energy is not linear. The formation energy of C is the highest and of Zr is close to zero explaining the good solubility of Zr in \hf until pure ZrO$_2$. Consequently, the formation energy of Ce which is the second lowest may indicate a similar good solubility in \hf. 

\begin{table}[tb]
\caption{\label{tab:formation_dopants} Shows the formation energy $E_f^\zeta [\text{D}^0]$ of the defects \donhf for \SI{3.125}{\formup} (\SI{6.25}{\formup}). The crystal radius r$_\text{c}$ is for the coordination number of the t-phase according to \cite{Shannon1976}.}
\begin{tabular}{cc|cccc}

D & r$_\text{c}$ & \multicolumn{4}{c}{$E_f^\zeta$ [D$^0_\text{Hf}$]} \\
 &  &  m & o & p-o & t \\
& pm &  \si{eV} & \si{eV} & \si{eV} & \si{eV} \\

\hline
C  &     29 &  9.1 (9.4) &  8.9 (9.2) &  9.1 (9.4) &  9.5 (-) \\
Si &     40 &  4.3 (4.0) &  4.2 (3.9) &  4.2 (3.6) &  2.8 (2.3) \\
Ge &     53 &  6.7 (6.5) &  6.8 (6.6) &  6.7 (6.6) &  5.7 (5.4) \\
Ti &     56 &  1.7 (1.7) &  1.8 (1.7) &  1.8 (1.8) &  1.6 (1.5) \\
Sn &     69 &  6.2 (6.2) &  6.2 (6.2) &  6.1 (6.1) &  6.0 (5.9) \\
Zr &     98 &  0.2 (0.2) &  0.2 (0.2) &  0.2 (0.2) &  0.1 (0.1) \\
Ce &    111 &  1.3 (1.3) &  1.2 (1.2) &  1.0 (1.0) &  0.7 (0.7) \\
\hline
\end{tabular}
\end{table}

In this section, total energy differences of the o-, p-o- and t-phase were presented and analyzed for the dopants C, Si, Ge, Ti, Sn, Zr and Ce concluding that none of the dopants alone promote the p-o-phase to the ground state. Besides silicon, which favours the p-o-phase but much more the t-phase, Ce is a promising candidate for doping since the t-phase is less favoured than with Si. Subsequently, the volume change due to the dopants are compared. Following the argumentation from Clima et al.\cite{Clima2014} that the volume is inversely proportional to the coercive field Ce doped \hf should expose ferroelectricity with a small coercive field. Finally, the formation energy of the dopants was investigated revealing that Ce has the second lowest formation energy in our comparison promoting a good solutibility. In addition, all attempts to find a general relation between a geometrical quantity and the energy differences or the formation energy failed. However, to gain an idea of the incorporation of the dopants into the host \hf crystal, the geometric neighborhood of the dopants is analyzed in the next section.

\subsection{Geometrical incorporation of the dopants}

To include chemical effects in the analysis we have evaluated the dopant to oxygen bond geometry. The bonding environments can be classified with polyhedrons. FIG. \ref{fig:bond_orbitals} shows the installation of silicon into the host crystal for (a) the t-phase bonding to the four neighbouring oxygen and (b) the p-o-phase bonding to the six neighbouring oxygen. In general, FIG. \ref{fig:bond_orbitals} exemplifies the incorporation of all the dopants into the host \hf crystal. To discriminate between oxygen neighbours with an active or inactive bond to the dopant and thus defining the coordination numbers $n$, we require the distance to be within the average bond length plus \SI{50}{pm}. This criterion matches very closely the average bond length as define by Baur \cite{Baur1974} in 1974. Instead of using Baur's fractional, effective coordination we use the integer coordination from counting.

For the different phases and dopants, TAB. \ref{tab:bondlength} collects the coordination number $n$ from the computed structures. For Ti, Sn and Zr the bond configuration is seven-fold and similar to undoped Hf, except for the t-phase. Zr has the same coordination as Hf itself confirming the chemical similarity. On the other hand, C, Si, Ge, Ti, and Sn are four-fold coordinated in the t-phase. In particular, C as the smallest dopant in this comparison differs in the bonding coordination significantly. In the less symmetric m-, o- and p-o-phase, C has only three bonds to oxygen suggesting that C left the substitutional position of Hf. Other more energetically favourable installations like interstitial or oxygen substitution of C in the crystal are possible but were not investigated in this study. Ge has six bonds in the three less symmetric phases. Si, being smaller, cannot build six bonds in the m- and o-phase, but only in the p-o-phase. Since the six bonds in the p-o-phase are stronger than the five bonds in the m- and o-phase, silicon may favour the p-o-phase relative to the other phases, with the exception of the t-phase. It seems that the six-fold coordination of silicon leads to the second strongest bond following the four-fold coordination. The special facilitation of the p-o-phase with silicon doping is a result of the adoption of the favourable six-fold coordination in comparison to the adoption of the unfavourable five-fold coordination in the competing m- and o-phases.

\begin{table}[tb]
\caption{\label{tab:bondlength} Calculated bond coordination $n$ and the distortion index $d$ for the four-valent dopants D. The crystal radius r$_\text{c}$ is for the coordination number $n$ of the t-phase according to \cite{Shannon1976}.}
\begin{tabular}{cc|cc|cc|cc|cc}
\hline
   &  &\multicolumn{2}{c|}{m} & \multicolumn{2}{c|}{o} & \multicolumn{2}{c|}{p-o} & \multicolumn{2}{c}{t}\\
D  &    r$_{c}$ & $n$ & $d$ & $n$ & $d$ & $n$ & $d$ & $n$ & $d$ \\
  & pm  &  & pm &  & pm &  & pm &  & pm \\
\hline
C  & 29  & 3 & 0.008  & 3 & 0.007 & 3 & 0.008 & 4 & 0.014 \\
Si & 40  & 5 & 0.017 & 5 & 0.030 & 6 & 0.036 & 4 & 0.000  \\
Ge & 53  & 6 & 0.034 & 6 & 0.033 & 6 & 0.061 & 4 & 0.000 \\
Ti & 56  & 7 & 0.055 & 7 & 0.057 & 7 & 0.046 & 4 & 0.000 \\
Sn & 69  & 7 & 0.038 & 7 & 0.031 & 7 & 0.034 & 4 & 0.000 \\
Hf & 97  & 7 & 0.026 & 7 & 0.026 & 7 & 0.021 & 8 & 0.062 \\
Zr & 98  & 7 & 0.026 & 7 & 0.027 & 7 & 0.022 & 8 & 0.059 \\
Ce & 111 & 7 & 0.028 & 7 & 0.027 & 7 & 0.028 & 8 & 0.026 \\
\hline
\end{tabular}
\end{table}

Along with the coordination number in TAB. \ref{tab:bondlength}, the distortion index $d$ is given. The distortion index describes the root mean square deviation of the bond length to the average bond length and is therefore a measure for the symmetry of the bond configuration\cite{Baur1974}. Smaller values indicate a better fit. 

\section{Conclusions}
We have explored the effect of silicon doping and other four-valued dopants on the crystallographic phase formation, especially of the p-o phase, in \hf from first principles. In a first step we evaluated different DFT methods -- LDA, PBE and HSE06 XC functionals, all-electron and plane wave -- for silicon doping and found all methods to predict qualitatively a strong stabilization of the t-phase and a weaker stabilization of the p-o-phase, such that the p-o-phase is below the t-phase only in a concentration window around \SIrange{3}{5}{\formup}. All methods agree that in this concentration window the m-phase and the o-phase are still lower revealing Si doping alone is insufficient to explain the favouring of the ferroelectric p-o-phase for monocrystalline material. Further mechanism for removing the m-phase and the o-phase from the ground state are required as has been discussed in previous work. An analysis of several possible defect states revealed that mainly the \sionhf defect is introduced from doping in ALD processes. Analyzing the the concentration dependence, nonlinear doping effects become visible which require a more thorough analysis of dopant-dopant interaction effects. To find possible systematic effects of \hf doping we calculated the effect of the four-valued dopants C, Ge, Ti, Sn, Ce and Zr on the phase stability. Besides Si, only Sn and Zr show a small stabilization effect of the p-o phase. The effect on the t-phase is known and was reproduced. The effects of doping on crystal volume are in the order of \SI{1}{\%}, but the related deformation energy turns out to be much smaller than the introduced formation energy such that the main effect of doping is chemical. The significant stabilization of the p-o phase with silicon turns out to be a very specific effect. As the promotion of the t-phase is related to the existence of a tetrahedral bonding configuration which is especially strong, the promotion of the p-o phase is related to the existence of a octahedral bonding configuration. For the other four-valued dopants, this bonding configuration either does not exist or, like in Ge, has this configuration in a very irregular shape. It is expected that the explanation of p-o-phase stabilization in \hf with other dopants like Al, Y, La, Gd has a different explanation.

Based on the calculations, Ce is a promising candidate for promoting ferroelectricity in \hf. The stabilization of the p-o-phase relative to the stabilization of the t-phase is good, promising a large window in concentration. Based on the small formation energy, the solubility in \hf is good and the volume increase with doping should lower the coercive field.


\begin{acknowledgement}
The author wants to thank U. Schr\"oder, T. Schenk, Min Hyuk Park from NamLab/SNU and U. B\"ottger, S. Starschich from RWTH Aachen for discussions. The German Research Foundation (Deutsche Forschungsgemeinschaft) is acknowledged for funding this research in the frame of the project ''Inferox'' (project no. MI 1247/11-1). The authors gratefully acknowledge the Gauss Centre for Supercomputing e.V. (www.gauss-centre.eu) for funding this project by providing computing time on the GCS Supercomputer SuperMUC at Leibniz Supercomputing Center (LRZ, www.lrz.de).
\end{acknowledgement}

\begin{suppinfo}
Lattice constants and band gaps for the different used DFT methods can be found in the following table:
\begin{table}
\caption{\label{tab:supp_lattice_and_gaps} Lattice constant and band gap for silicon doped \hf.}
\begin{tabular}{ll|ccccc}
\hline
Phase   &   & \multicolumn{3}{c}{AIMS}  & \multicolumn{2}{c}{ABINIT} \\
        &   & LDA & PBE & HSE06 & GBRV & GBRV*  \\
\hline
\multirow{4}{*}{m} & band gap in eV &     3.92 &     4.13 &       5.73 &        - &         4.31 \\
  & a in \AA &    10.07 &    10.28 &      10.28 &       10.04 &        10.11 \\
  & b in \AA &     5.08 &     5.16 &       5.16 &        5.07 &         5.10 \\
  & c in \AA &    10.51 &    10.75 &      10.75 &       10.46 &        10.53 \\ \hline
\multirow{4}{*}{o} & band gap in eV &     4.00 &     4.07 &       - &        - &         - \\
  & a in \AA &     9.99 &    10.20 &       - &        - &         - \\
  & b in \AA &    10.17 &    10.33 &       - &        - &         - \\
  & c in \AA &     5.22 &     5.33 &       - &        - &         - \\ \hline
\multirow{4}{*}{p-o} & band gap in eV &     4.15 &     4.29 &       5.94 &        - &         4.31 \\
  & a in \AA &     9.92 &    10.10 &      10.10 &        9.89 &         9.96 \\
  & b in \AA &     5.18 &     5.31 &       5.31 &        5.17 &         5.20 \\
  & c in \AA &     9.93 &    10.13 &      10.13 &        9.90 &         9.96 \\ \hline
\multirow{4}{*}{t} & band gap in eV &     4.22 &     4.59 &       5.73 &        4.31 &         4.31 \\
  & a in \AA &    10.05 &    10.16 &      10.75 &       10.00 &        10.06 \\
  & b in \AA &     4.99 &    10.57 &       5.16 &        4.98 &         5.01 \\
  & c in \AA &    10.05 &     5.13 &      10.28 &       10.00 &        10.06 \\

\hline
\end{tabular}
\end{table}

\begin{figure}[tbp]
\includegraphics[width=0.9\textwidth]{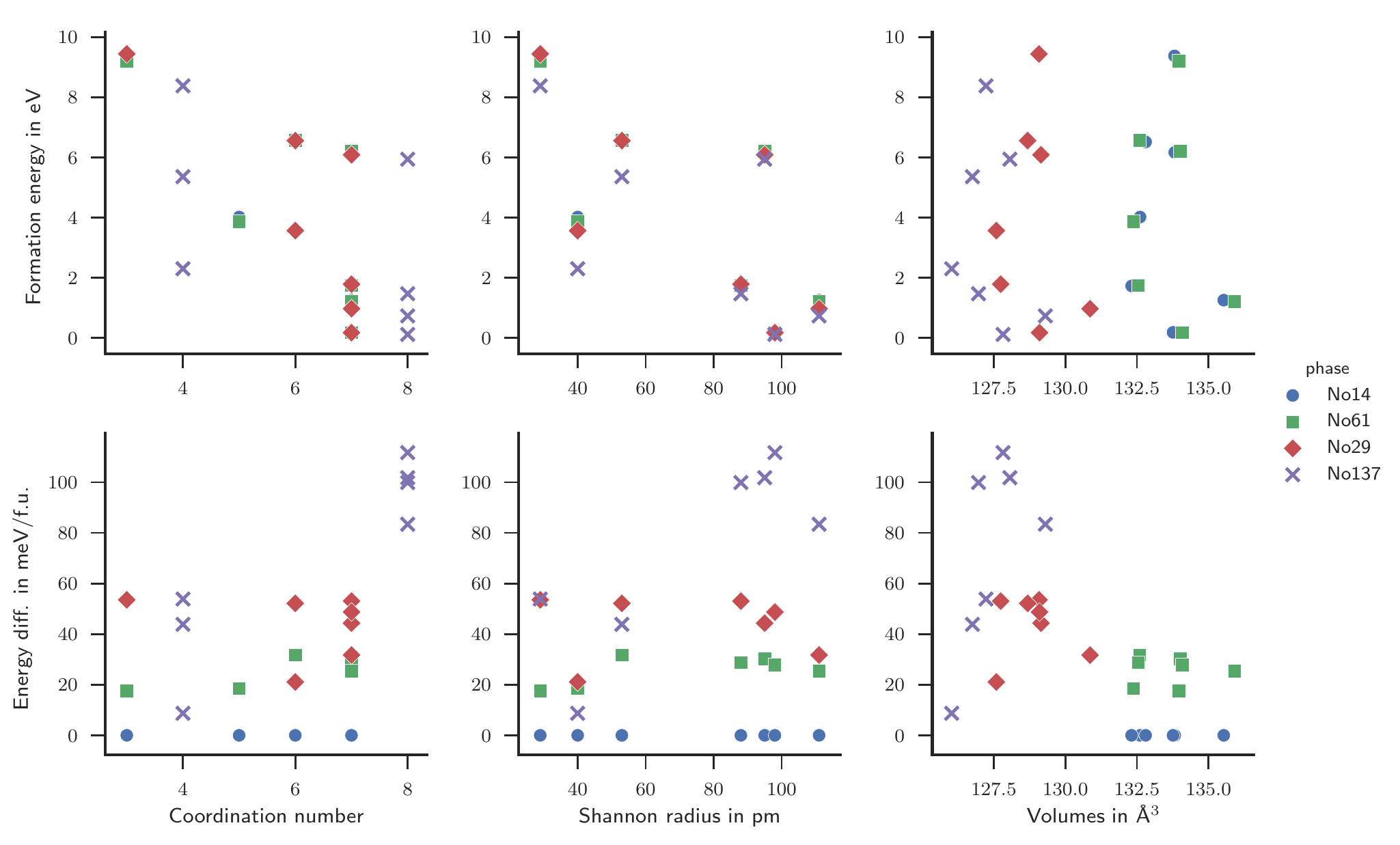}
\caption{\label{fig:croo_3} Different correlations of \SI{6.25}{\formup} doping. }
\end{figure}

\begin{figure}[tbp]
\includegraphics[width=0.9\textwidth]{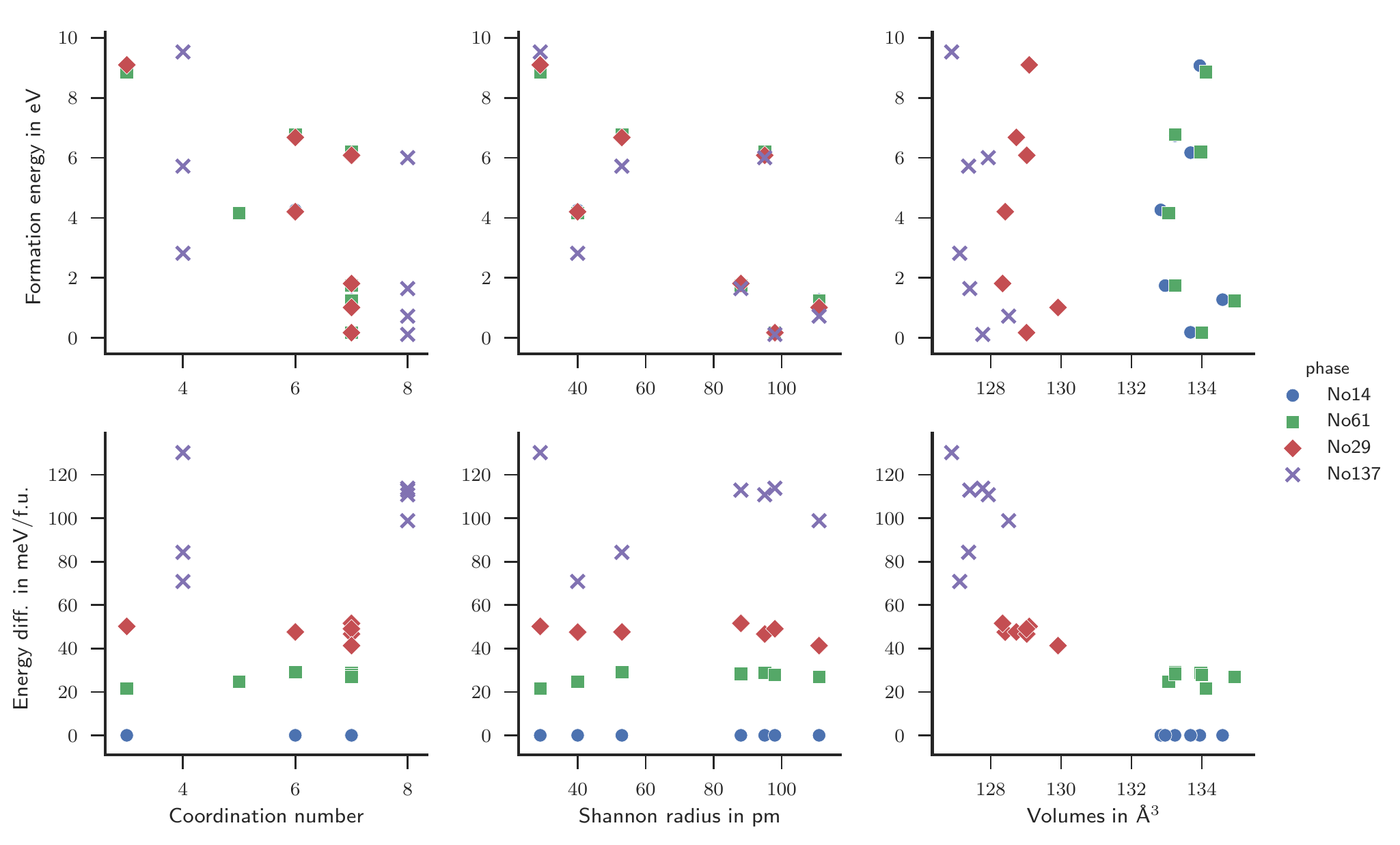}
\caption{\label{fig:croo_3} Different correlations of \SI{3.125}{\formup} doping. }
\end{figure}

\begin{figure}[tbp]
\includegraphics[width=0.5\textwidth]{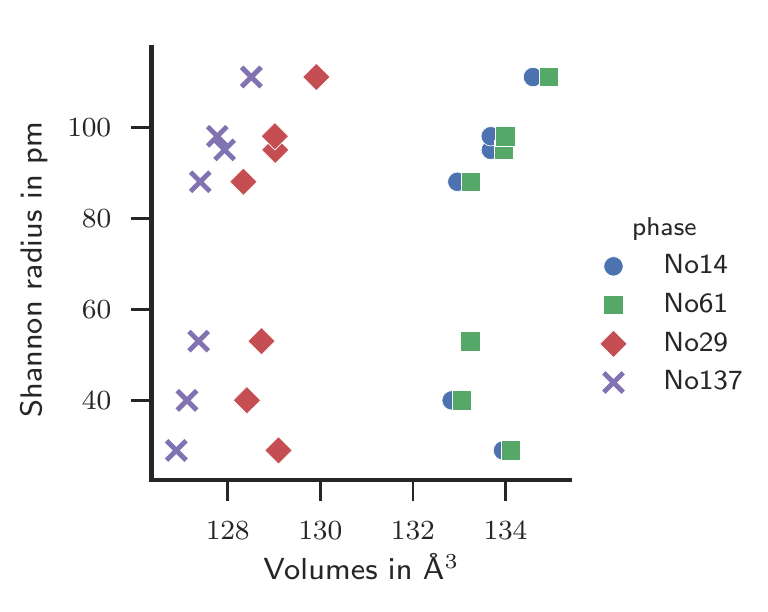}
\caption{\label{fig:croo_3} Volume vs. Shannon radius for \SI{3.125}{\formup} doping. }
\end{figure}

\end{suppinfo}

\bibliography{bib}

\end{document}